\documentclass [aps,showpacs,showkeys,preprint,amsmath,amssymb]{revtex4}
\usepackage{txfonts}
\usepackage{graphicx}
\usepackage{amsmath}

\begin{document}

\title{ \bf Interfacial dead layer effects on current-voltage characteristics in asymmetric ferroelectric tunnel junctions }
\author{\bf  Ping Sun$^{1,2}$, Yin-Zhong Wu$^{1,*}$, Su-Hua Zhu$^{1}$, Tian-Yi Cai$^{2}$, and Sheng Ju$^{2}$}
\affiliation{$^{1}$ Jiangsu Laboratory of Advanced Functional Materials and Physics Department, Changshu Institute of Technology, Changshu 215500, China\footnote{Corresponding author\\ Email address: yzwu@cslg.edu.cn}\\
$^{2}$School of Physical Science and Technology,Soochow University, Suzhou 215006, China}

\vspace{20cm}

\begin{abstract}
Current-voltage characteristics and $P-E$ loops are simulated in SrRuO$_{3}$/BaTiO$_{3}$/Pt tunneling junctions with interfacial dead layer. The
unswitchable interfacial polarization is coupled with the screen charge and the barrier polarization self-consistently within the Thomas-Fermi
model and the Landau-Devonshire theory. The shift of $P-E$ loop from the center position and the unequal values of the positive coercive field
and the negative coercive field are found, which are induced by the asymmetricity of interface dipoles. A complete $J-V$ curve of the junction
is shown for different barrier thickness, and the effect of the magnitude of interfacial polarization on the tunneling current is also
investigated.
\end{abstract}

\pacs{73.40.Gk, 77.55.fe, 77.80.bn} \keywords{Current-voltage characteristics; Interfacial dead layer; Ferroelectric tunneling junction;}

\maketitle
\section{Introduction}
With the development of fabrication technologies, ultrathin ferroelectric films used as tunneling barriers have attracted significant interest
in many fields\cite{Science}. Typically, ferroelectric tunnel junction(FTJ) prepared by using an epitaxial ultrathin ferroelectric layer as a
tunnel barrier sandwiched between two mental electrodes has brought great progress in experiment and theory \cite{nature1,nano lett1}. In the
past decade, it was shown that ferroelectricity could be maintained in perovskite oxides films with thickness of the order of a few
nanometers\cite{APL1,PRL1}, these experimental results were consistent with first-principles calculations predicting the critical thickness of
ferroelectric barrier could as thin as a few lattice parameters\cite{PRL2}, and the existence of ferroelectricity in ultrathin films gives
possibilities for nonvolatile memories, such as ferroelectric and muliferroic tunnel junctions. Compared with conventional memories,
ferroelectric memories have the advantage of high read and write speed and high density data storage\cite{nature1}. The existed researches
confirm that the reversal of the electric polarization in ferroelectric barrier produces a change in the electrostatic potential profile across
the junction, and this leads to the resistance change which can reach a few orders of magnitude for asymmetric metal electrodes\cite{PRL3}. As
is well known, interfaces exist inevitably in FTJs, and the interface will give rise to great influence on the transport properties of FTJs.
Glinchuk and Morozovska\cite{JPD} introduced an interfacial dipole between the ferroelectric thin film and its substrate, and they claimed that
the interfacial dipole was originated from the mismatch between the lattice constants and thermal coefficients of the film and its substrate as
well as growth imperfections. Duan\cite{nano lett2} carried out first-principles calculations of KNbO$_{3}$ thin film placed between two metal
electrodes, they found that bonding between the metal and ferroelectric atoms at the interface induced an interfacial dipole moment, which is
electrode dependent. Liu\cite{PRB1} demonstrated that a BaO/RuO$_{2}$ interface in SrRuO$_{3}$/BaTiO$_{3}$/SrRuO$_{3}$ epitaxial heterostructure
grown on SrTiO$_{3}$(STO) can lead to a nonswitchable polarization state for thin BaTiO$_{3}$(BTO) films due to a fixed interfacial dipole, our
group also reported that tunneling electroresistance(TER) can be induced by the asymmetric interfaces in a FTJ with symmetric
electrodes\cite{JAP1}. However, previous studies are focused on the effect of interface on the polarization of ferroelectric barrier, and their
efforts are mostly concentrated on zero bias conductance of FTJs, and study of the interfacial effect on the current-voltage characteristics is
rare. Natalya A. Zimbovskaya\cite{JAP2} and the group of Wenwu Cao\cite{JAP3,JAP4} had carried out investigations on the current-voltage
characteristics of FTJs, but the magnitude and the direction of the barrier's polarization in their models are given artificially and
externally, and the coercive field has not been given within their theoretical studies. In this paper, the typical SrRuO$_{3}$/BaTiO$_{3}$/Pt
(SRO/BTO/Pt) structure is selected as an example to investigate the transport property of an asymmetric FTJ within the framework of
Landau-Devonshire theory and quantum tunneling theory. SrRuO$_{3}$(SRO) is a suitable oxide electrode for the epitaxial growth due to its
similar lattice constant with the substrate and BTO, and Pt is a well conductive metal electrode. The interfacial polarization, inhomogeneous
barrier polarization and the screen charge of the electrodes are coupled together in our model, and the coercive field for the switching of
barrier polarization is obtained self-consistently. It is found that the interfacial dead layers will have great influence on the $P-E$
hysteresis loop and $J-V$ behavior of FTJs, and shifts of the $P-E$ hysteresis loop and the $J-V$ loop are observed. The effects of the
magnitude of interfacial polarization on the tunneling current is also studied, we hope our investigations will bring useful guidance in the
studying of the interfacial effects in nanoscale devices.

\section{Model and Theory}
The typical SRO/BTO/Pt junction(See Fig.~1) is used to simulate the $P-E$ and $J-V$ hysteretic behavior of the ferroelectric junction. It is
assumed that the polarization of the BTO barrier is orthogonal to the electrode, which can be realized through the misfit compression stress
from the substrate. In Fig.~1, P$_{iL}$ and P$_{iR}$ are the polarization of the left and right interface, respectively, and P$_{iL}$(P$_{iR}$ )
is assumed to be fixed\cite{nano lett2}. From the first-principles calculations, it is found that the effects of the interfacial dead layer
extend for only 1-2 atomic layers\cite{PRB2}. Therefore, the thickness of the interface layer is chosen as one unit cell in this paper, i.e.,
d$_{iL}$=d$_{iR}$=4\AA . Based on the lattice model for a strained nanoscale ferroelectric capacitor\cite{APL2}, the average free-energy density
of the BTO barrier can be written as

\begin{eqnarray}
 F &=& \frac{1}{n}\bigg\{\sum\limits_{i=1}^n[\alpha^*_{1}P^2_{i}+\alpha^*_{11}P^4_{i}+\alpha_{111}P^6_{i}-\frac{1}{2}E^i_{d}P_{i}-E_{ext}P_{i}]\nonumber\\
   & &
   {}+\sum\limits_{i=2}^n\frac{G_{11}}{2}(\frac{P_{i}-P_{i-1}}{c})^2+\frac{1}{2}\frac{G_{11}}{2}(\frac{P_{1}-P_{iL}}{\delta_{L}})^2+\frac{1}{2}\frac{G_{11}}{2}(\frac{P_{n}-P_{iR}}{\delta_{R}})^2\bigg\},
\end{eqnarray}
where $\alpha^*_{1}=\alpha_{1}-\frac{2U_{m}Q_{12}}{S_{11}+S_{12}}$, $\alpha^*_{11}=\alpha_{11}+\frac{Q^2_{12}}{S_{11}+S_{12}}$, $\alpha_{1}$ is
the dielectric stiffness coefficient at constant strain, $S_{11}$, $S_{12}$  are the elastic compliance, $Q_{12}$ is the electrostricitive
coefficient, $G_{11}$ is the coefficient of the gradient terms in the free-energy expansion, and $U_{m}=(a_{STO}-a_{BTO})/a_{STO}$ is the
in-plain strain, the tetragonal BTO barrier consists of atomic dipole moments $p_i$ orthogonal to the electrode with an infinite extension along
the $y$ and $z$ axis, the barrier can be divided into $n$ layers along $x$ axis, where the thickness of each layer is a unit cell. The first
summation in Eq.~(1) is carried out over all the layers within the barrier. The $ith$ layer polarization P$_i$ is given by $P_{i}=p_{i}/abc$,
where $a$, $b$, $c$ are the lattice constant along $y$-axis, $z$-axis, and $x$-axis, respectively. The first three terms in the Eq.~(1) are the
standard Landau-Devonshire energy density, $\sum\limits_{i=1}^n-\frac{1}{2}E^i_{d}P_{i}$ represents the self-electrostatic energy density,
$E^i_{d}$ is the depolarization field, $-\sum\limits_{i=1}^nE_{ext}P_{i}$ denotes an additional electrical energy density that is due to an
external applied electric field, $\sum\limits_{i=2}^n\frac{G_{11}}{2}(\frac{P_{i}-P_{i-1}}{c})^2$ denotes the gradient energy, which represents
the inhomogeneous polarization's contribution to the free energy density. The last two terms denote the interface contribution to the
free-energy\cite{nano lett2}, $c$ is the lattice constant along $x$ axis, and $\delta_{L}$($\delta_{R}$) is extrapolation length which
represents the discrepancy between the interface and the interior of the thin film.
\begin{figure}
\vskip -0.9cm
\includegraphics[width=0.75\textwidth]{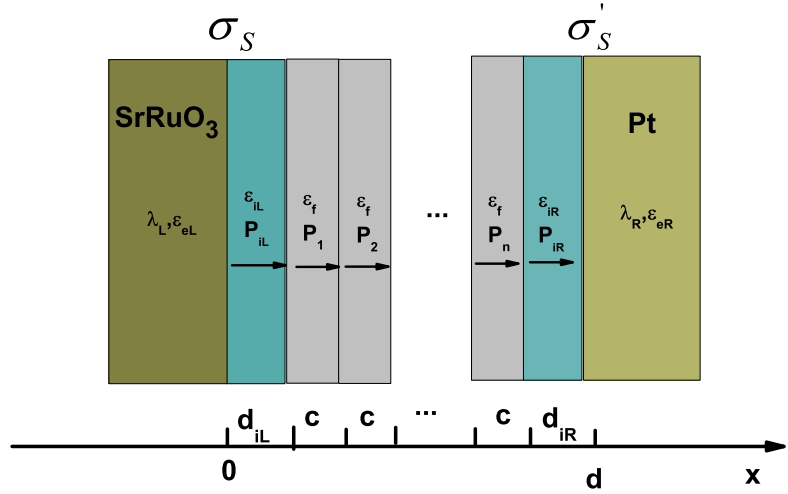}
\vskip -0.5cm
  \caption{\it{Sketch of the asymmetric FTJ with two interfacial dead layers,
   $\sigma_{s}$  and  $\sigma'_{s}$  stand for the screening charges in the left and right electrodes, respectively.}} \vskip 0.6cm
\end{figure}

According to the charge conservation, the magnitude of screening
charge density $\sigma_{s}$ in two electrodes is the same. Under the
framework of Thomas-Fermi model, the screening potentials within the
left and right electrodes can be written as\cite{BOOK1}
\begin{equation}
  \varphi(x)=
  \Bigg\{\begin{array}{lc}
   \frac{\sigma_{s}\lambda_{L}}{\varepsilon_{L}}e^{-|x|/\lambda_{L}}, \hskip 1 cm & x\leq 0,\\
   -\frac{\sigma_{s}\lambda_{R}}{\epsilon_{L}}e^{-|x-d|/\lambda_{R}}, \hskip 1 cm & x\geq d,\\
  \end{array}
\end{equation}
where $\lambda_{L}$($\lambda_{R}$) is the screening length in the left (right) electrode, $\epsilon_{L}$($\epsilon_{R}$) is the dielectric
constant of the left (right) electrode, and $d$ is the thickness of the total barrier including the interface layer. Using the continuity
conditions of electric displacement vectors and electrostatic potential at the boundaries, we obtain

\begin{equation}
\sigma_{s}=
\frac{V+\frac{P_{iL}d_{iL}}{\epsilon_{iL}}+\frac{\overline{P_{f}}d_{f}}{\epsilon_{f}}+\frac{P_{iR}d_{iR}}{\epsilon_{iR}}}
{\frac{\lambda_{L}}{\epsilon_{L}}+\frac{d_{iL}}{\epsilon_{iL}}+\frac{d_{f}}{\epsilon_{f}}+\frac{d_{iR}}{\epsilon_{iR}}
+\frac{\lambda_{R}}{\epsilon_{R}}},
\end{equation}
where $\epsilon_{iL}$ and $\epsilon_{iR}$ are the dielectric constants of the left and right interface, respectively. $\overline{P_{f}}$, $d_f$,
and $\epsilon_{f}$ are the average spontaneous polarization, the thickness, and the dielectric constant of the BTO barrier. The depolarization
fields within the interfaces and each ferroelectric layer have the following form:
\begin{equation}
 \begin{array}{cccc}
 E_{iL}^{d}&=&\frac{\sigma_{s}-P_{iL}} {\epsilon_{iL}},\\
 E_{i}^{d}&=&\frac{\sigma_{s}-P_{i}}{\epsilon_{f}},\\
 E_{iR}^{d}&=&\frac{\sigma_{s}-P_{iR}}{\epsilon_{iR}}.\\
 \end{array}
\end{equation}
The screening lengths for the left and right electrodes are selected
as $\lambda_{L}$=0.8\AA \ and $\lambda_{R}$=0.4\AA, and the
corresponding dielectric constants are taken as
$\epsilon_{L}$=8.85$\epsilon_{0}$  and
$\epsilon_{R}$=2$\epsilon_{0}$, these parameters for the left and
the right electrodes are typical values for SRO and Pt
electrodes\cite{PRL4}, respectively. The magnitude of the interface
dielectric constant take the average value
$\epsilon_{iL}=(\epsilon_{L}+\epsilon_{f})/2$ and
$\epsilon_{iR}=(\epsilon_{R}+\epsilon_{f})/2$.

The polarization $P_{i}$ in thermodynamic equilibrium state can be derived by the equations $\partial F/\partial P_{i}=0(i=1, 2, \cdots, n)$ and
boundary conditions,

\begin{equation}
  2\alpha^*_{1}P_{i}+4\alpha^*_{11}P^3_{i}+6\alpha_{111}P^5_{i}-\frac{1}{2}\frac{\partial }{\partial
  P_{i}}(E^i_{d}P_{i})+G_{11}\frac{P_{i}-P_{i-1}}{c^2}-G_{11}\frac{P_{i+1}-P_{i}}{c^2}-E_{ext}=0,
\end{equation}

\begin{equation}
  (P_{1}-\delta_{L}\frac{dP}{dx})_{x=0}=P_{iL},\
  (P_{n}+\delta_{R}\frac{dP}{dx})_{x=d}=P_{iR}.
\end{equation}
The detail coefficients in Landau free energy in Eq.~(5), lattice
constants, and the dielectric constants for BTO are listed in
Ref.~\cite{para}.  As is well known, due to piezoelectric effect,
the effective barrier thickness, effect electron mass, and barrier
conduction band edge in a FTJ will be changed under an applied
electric field. Cao group\cite{JAP4} verified that the
depolarization effect is much greater than the piezoelectric effect.
Therefore, to avoid extra complications in further computations, the
piezoelectric effect is neglected in our following calculations.
Within the framework of Landau theory, the microscopic profile of
polarization in each ferroelectric layer with consideration of the
interfacial dead layer can be calculated numerically, therefore, the
electrostatic profile $\varphi$(x) is given as
\begin{equation}
   \varphi(x)=\left\{
   \begin{array}{lc}
   \frac{\sigma_{s}\lambda_{L}}{\epsilon_{L}}-E_{iL}^{d}\cdot x, & 0< x\leq d_{iL},\\
   \frac{\sigma_{s}\lambda_{L}}{\epsilon_{L}}-E_{iL}^{d}\cdot d_{iL}-E_{1}^{d}\cdot(x-d_{iL}), & d_{iL}< x \leq d_{iL}+c,\\
   \frac{\sigma_{s}\lambda_{L}}{\epsilon_{L}}-E_{iL}^{d}\cdot d_{iL}-E_{1}^{d}\cdot c-E_{2}^{d}\cdot(x-d_{iL}-c), & d_{iL}+c< x \leq d_{iL}+2c,\\
              \cdots                                                       & \cdots                     \\
   \frac{\sigma_{s}\lambda_{L}}{\epsilon_{L}}-E_{iL}^{d}\cdot d_{iL}-E_{1}^{d}\cdot c-E_{2}^{d}\cdot c-\cdots-E_{iR}^{d}\cdot(x-d_{iL}-nc),& d-d_{iR}< x\leq d.
   \end{array}\right.
\end{equation}
The overall potential profile $U(x)$ across the junction is the superposition of the electrostatic energy potential -e$\varphi$(x), the
electronic potential in the electrodes, the rectangular potential $U_{iL}$($U_{iR}$) in the left(right) interface, and $U_{f}$ in FE barrier.
Based on the potential energy distribution, the tunneling current through the junction can be calculated, the current per unit area $J$ can be
derived form the following formula\cite{JAP2}

\begin{equation}
\emph{J}=\frac{2e}{h}\int\\
dE[f_{L}-f_{R}]\int{\frac{d^2k_{\parallel}}{(2\pi)^{2}}T(E_{F},k_{\parallel})},
\end{equation}
where $ T(E_{F},k_{\parallel})$ is the transmission coefficient at the Fermi energy for a given transverse wave vector $k_{\parallel}$, the
transmission coefficient is obtained by solving numerically the Schr$\ddot{o}$dinger equation for an electron moving in the total potential
$U(x)$ by imposing a boundary condition of the incident plane wave and by calculating the amplitude of the transmitted plane wave within the
formation of transfer matrix, and $f_{L,R}$ are Fermi-Dirac distribution functions with chemical potentials $\mu_{L,R}$, when a nonzero bias
voltage \emph{V} is applied across the junction, $\mu_{L}$ and $\mu_{R}$ satisfy $\mu_{R}-\mu_{L}=eV$, and the Fermi-Dirac distribution
functions are expressed as $f_{L}(E_{L})=\{1+e^\frac{E_{L}-\mu_{L}}{k_{B}T}\}^{-1}$ and $f_{R}(E_{R})=\{1+e^\frac{E_{R}-\mu_{R}}{k_{B}T}\}^{-1}$
, where $k_{B}$ is the Boltzmann constant. We also assume the electron has a free electron mass, the Fermi energy is chosen as $E_{F}=3.0eV$,
and $U_{iL}=U_{iR}=0.6eV$ and $U_{f}=0.6eV$ in the interfaces and the barrier\cite{PRL3}.

\section{ Results and Discussions}
$P$-$E$ hysteresis loops of the SRO/BTO/Pt junction are plotted in Fig.~2 and Fig.~3. For simplicity, we use $P$ to stand for $\overline{P_f}$
in this paper, which is the average polarization of the ferroelectric barrier. In SrRuO$_{3}$/KNbO$_{3}$/SrRuO$_{3}$ and Pt/KNbO$_{3}$/Pt
junctions studied in Ref.~\cite{nano lett2}, the interface dipole moment points to the barrier for SRO electrodes, while the interface dipole
moment is pointing to the electrodes for Pt electrodes. Based on first-principles calculations, the BaO/RuO$_{2}$ interface dipole is
nonswitchable and points to the barrier BTO\cite{PRB1}. Therefore, the interface polarizations are assumed to be fixed, and take the value of
$P_{iL}=0.1C/m^2$ and $P_{iR}=0.2C/m^2$ in studying the $P-E$ hysteresis behavior of SRO/BTO/Pt junctions in this paper, and the positive values
of $P_{iL}$ and $P_{iR}$ imply that they are always pointing to the right, as shown in Fig.~1. For a given applied electric field $E$, each
$P_{i}$ can be numerically obtained from Eq.~(5). Then, the averaged barrier polarization $P$ as a function of $E_{ext}$ is plotted, and the
value of coercive field can be found from the P-E loop. The profile of the total potential energy is achieved when each $P_{i}$ is given, and
the tunneling coefficient and consequently the tunneling current can be obtained from Eq.~(8). Therefore, the coercive field in the $J-V$ curve
in our system is consistent with the ferroelectric switching. From Fig.~2, one can see that a symmetrical $P-E$ loop occurs without
consideration of the interfacial layer. The consideration of the interfacial layer will cause a shift of the $P-E$ loop, and the shift takes
place along the direction of positive external field. The shift of $P-E$ loop is caused by an intrinsic bias field, which is induced by
asymmetric interfacial dipoles, and the direction of the intrinsic bias field is opposite to the external field, therefore, $P-E$ loop will move
to the positive direction of the applied field. The offset of $P-E$ loop on the thickness of the barrier is shown in Fig.~3. One can see, from
Fig.~3, that the offset increases with the decrease of the barrier thickness under identical interfacial dipoles. As the shift of $P-E$ loop
originates from the asymmetric interfacial dipoles, the effect of the interface layer on the hysteresis behavior will become more remarkable for
a thinner barrier, so does the offset of $P-E$ loop for the ferroelectric junction with a thinner barrier.
\begin{figure}
\vskip -0.5cm
\includegraphics[width=0.65\textwidth]{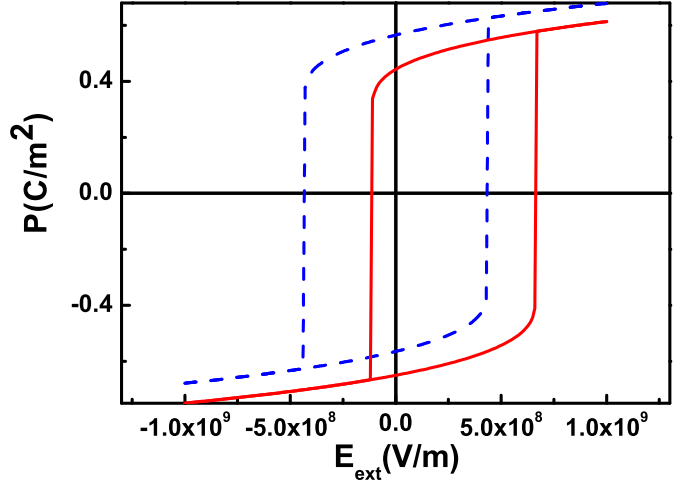}
\vskip -0.6cm
  \caption{\it{$P-E$ hysteresis loops of the SRO/BTO/Pt junction with interface layers(solid line)
  and without interface layers (dashed line). The barrier thickness
  is 2.4 nm.}} \vskip -0.6cm \vskip 0.8cm
\end{figure}
\begin{figure}
\vskip -0.5cm
\includegraphics[width=0.65\textwidth]{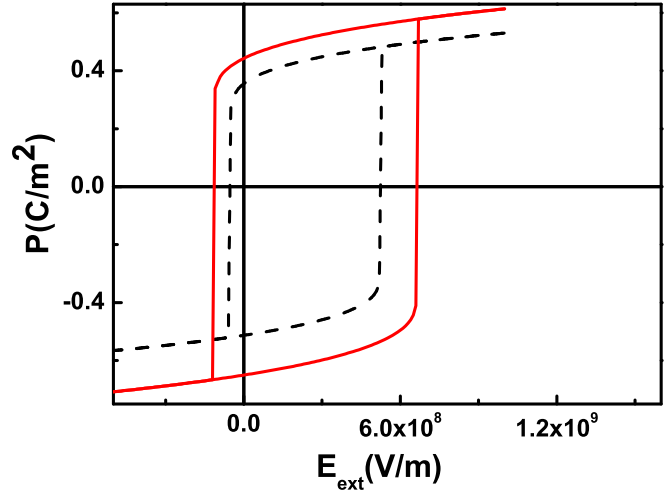}
\vskip -0.6cm
  \caption{\it{$P-E$ loops of the SRO/BTO/Pt junction with interface layers
   for different thickness of the BTO barrier: 1.6 nm(dashed line) and 2.4 nm(solid line).}} \vskip -0.6cm
\vskip 0.8cm
\end{figure}

Tunneling currents of SRO/BTO/Pt junctions can be calculated through Eq.~(8). The thickness of BTO barrier is selected as 1.6 $nm$, and the
complete $J-V$ curves are shown in Fig.~4. The dashed lines in Fig.~4 stand for the increasing direction of the scanning voltage, and solid
lines correspond to the opposite direction of the scanning voltage. Fig.~4(a) shows the $J-V$ curves for the ferroelectric junction without
consideration of interfacial dead layers, and equivalent magnitude of the positive and negative coercive field are obtained at $V^+_{C}=0.52V$
and $V^-_{C}=-0.52V$. With the consideration of interfacial dead layer in Fig.~4(b) (the parameters of the interfacial polarization are the same
as those in Fig.~3), the current shows jumps at $V^+_{C}=1.2V$ and $V^-_{C}=-0.1V$, which correspond to the switching of the average spontaneous
polarization of BTO. The asymmetry between the positive coercive field and the negative coercive field in $J-V$ curve is also induced by the
asymmetric interfacial dipoles, and is similar to the phenomenon mentioned above in the $P-E$ hysteresis behavior.
\begin{figure}
\vskip -0.6cm
\includegraphics[width=1.0\textwidth]{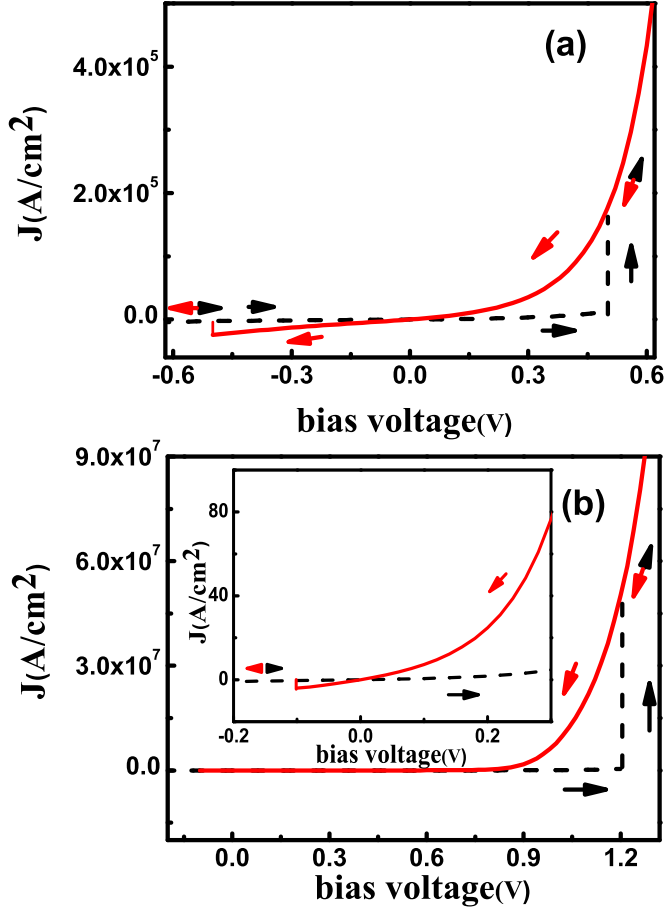}
\vskip -0.5cm
  \caption{\it{ Current-voltage characteristic of SRO/BTO/Pt junction (a) without consideration of interfacial
  layers, and (b) with interfacial dead layers. The barrier thickness of BTO is 1.6nm,
arrows indicate scanning directions of the applied voltage, and the
inset in (b) is an enlarged image corresponding to the low range of
the bias voltage.}} \vskip -0.6cm \vskip 0.7cm
\end{figure}

The current density as a function of the magnitude of the interfacial polarization is given in Fig.~5 for a fixed bias voltage 1.0 $V$. Dashed
lines correspond to the case for the barrier thickness 1.6 $nm$, and solid lines represent the case for 2.4 $nm$ BTO barrier. Compared Fig.~5(a)
with 5(b), it is found that $J$ increases with the increase of the magnitude of the left interfacial polarization $P_{iL}$, while $J$ decreases
with the increase of $P_{iR}$. The reason is that the direction of $P_{iL}$ is pointing to the barrier, and the increase of $P_{iL}$ will result
in the decrease of the average height of the barrier(See inset in Fig.~5a), while the increase of  $P_{iR}$ will give rise an increase of the
averaged barrier potential because $P_{iR}$ is pointing away from the barrier(See inset in Fig.~5b). This behavior is analogous to the TER
effect in ferroelectric junctions, which has been studied extensively in our previous work\cite{JAP6}. One can also see, from Fig.~5, that the
current density $J$ changes more quickly for the junction with a thinner barrier. This is due to the effect of interface on the transport
property becomes more remarkable with the increase of the proportion of interface in ferroelectric junctions with thinner barrier.

\begin{figure}
\vskip -0.9cm
\includegraphics[width=1.0\textwidth]{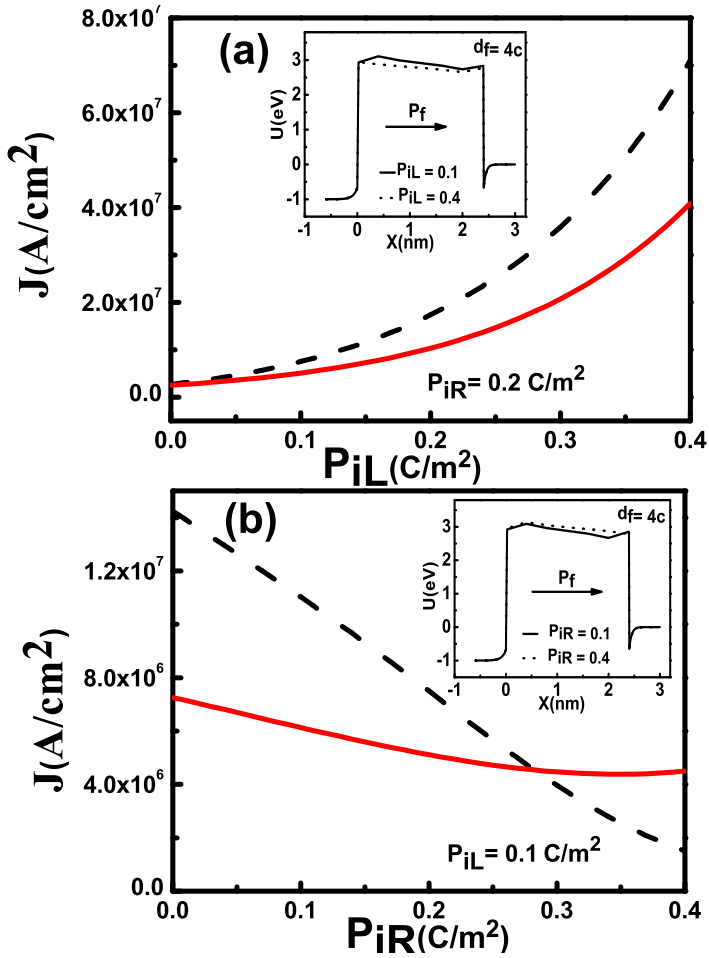}
\vskip -0.2cm
  \caption{\it{Current density as a function of (a) the magnitude of the left interface polarization $P_{iL}$ and (b) the magnitude of
$P_{iR}$  for different barrier thickness: 1.6 nm(dashed line) and 2.4 nm(solid line). Here, the bias voltage is taken as 1.0 $V$. The insets
are the profile of total potential energy for different values of $P_{iL}$ and $P_{iR}$.}}
 \vskip 0.6cm
\end{figure}

In summary, we have investigated the $P-E$ behavior and current-voltage characteristics of the asymmetric ferroelectric tunnel junction with the
interfacial dead layer, the inhomogeneous barrier polarization, the interfacial polarization and the screening charge of the electrodes are
coupled together in our model, shift of the $P-E$ hysteresis loop and the $J-V$ curve are observed. Effects of the magnitude of the interfacial
polarization on the tunneling current are also investigated. As is well known, the structure and property of the interface is very difficult to
manipulate and detect. To a certain extent, our model is simply, a more practical thickness, dielectric constant, and polarization of the
interface layer are needed to give more reliable simulations. The interface effect of a general FTJ besides the SRO/BTO/Pt junction will be
investigated in the future.

\begin{acknowledgments}
This work was supported by the National Science Foundation of
China(Grant Nos.11274054, 11047007, 11104193 and 10974140), the
QinLan project of Jiangsu Provincial Education Committee, and the
open project of Jiangsu Laboratory of Advanced Functional
Materials(12KFJJ005).
\end{acknowledgments}

\newpage

Figures' Caption\\

FIG.1. Sketch of the asymmetric FTJ with two interfacial
dead layers, $\sigma_{s}$  and  $\sigma'_{s}$  stand for the screening charges in the left and right electrodes, respectively.\\

FIG.2. $P-E$ hysteresis loops of the SRO/BTO/Pt junction with
interface layers(solid line) and without interface layers (dashed
line). The barrier thickness is 2.4 nm.\\

FIG.3. $P-E$ loops of the SRO/BTO/Pt junction with
interface layers for different thickness of the BTO barrier: 1.6 nm(dashed line) and 2.4 nm(solid line).\\

FIG.4. Current-voltage characteristic of SRO/BTO/Pt junction (a)
without consideration of interfacial dead
  layers, and (b) with interfacial dead layers. The barrier thickness of BTO is 1.6nm,
arrows indicate scanning directions of the applied voltage, and the
inset in (b) is an enlarged image corresponding to the low range
of the bias voltage.\\

FIG.5. Current density as a function of (a) the magnitude of the left interface polarization $P_{iL}$ and (b) the magnitude of $P_{iR}$  for
different barrier thickness: 1.6 nm(dashed line) and 2.4 nm(solid line). Here, the bias voltage is taken as 1.0 \emph{V}. The insets
are the profile of total potential energy for different values of $P_{iL}$ and $P_{iR}$.\\

\end{document}